\documentclass[prl,twocolumn,showpacs,letter]{revtex4-1}% 

\usepackage{amsmath,amsfonts}
\usepackage{graphicx}
\usepackage{amsmath,amssymb,psfrag,textcomp}
\usepackage{amsthm,upgreek,bm}

\newcommand{\bI}{\ensuremath{\mathbf{I}}}
\newcommand{\bG}{\ensuremath{\mathbf{G}}}
\newcommand{\bJ}{\ensuremath{\mathbf{J}}}
\newcommand{\bS}{\ensuremath{\mathbf{S}}}
\newcommand{\bH}{\ensuremath{\mathbf{H}}}
\newcommand{\bL}{\ensuremath{\mathbf{L}}}
\begin{document}
\title{Spectra of sparse regular graphs with loops}
\author{F. L. Metz$^{1}$, I. Neri$^{2,3}$, and D. Boll\'e$^{1}$} 
\affiliation{$^{1}$Instituut voor Theoretische Fysica, Katholieke 
Universiteit Leuven, Celestijnenlaan 200D, B-3001
Leuven, Belgium \\ 
${^2}$Universit\'e Montpellier 2, Laboratoire Charles Coulomb UMR 5221,
F-34095, Montpellier, France\\
${^3}$CNRS, Laboratoire Charles Coulomb UMR 5221, F-34095, Montpellier,
France
}

\begin{abstract}
We derive exact equations that determine the spectra of undirected and directed
sparsely connected regular graphs containing loops
of arbitrary length. The implications
of our results to the structural and dynamical
properties of networks are discussed by showing how loops influence
the size of the spectral gap
and the propensity for synchronization.
Analytical formulas
for the spectrum are obtained for specific
length of the loops.
\end{abstract}
\maketitle

Networks have emerged as a unified framework to study complex
problems in disciplines ranging from physics, biology, information theory,
chemistry to technological and social sciences \cite{Bar}.  
Some notable
examples are the backbone of the Internet, which consists of routers connected
by physical links, and the metabolism of the cell, represented as a tripartite
network of metabolites, reactions and enzymes. 
As many seemingly unrelated
problems are modeled by networks, it is crucial to
understand how the topology of networks influences the 
processes governed on them.  The efficiency of
error-correcting codes
and communication networks \cite{Murr, Hoory}, the propensity for 
synchronization \cite{Pec, Don} and the mixing 
times of search algorithms \cite{Lov}, among others, are
unveiled from a spectral analysis, i.e.~from a study of the adjacency matrix and
the Laplacian of the
network \cite{Chung}.

A widespread theoretical approach consists in modeling real-world
networks by sparsely connected 
random graphs \cite{Bol}, which have
 a local tree-like structure and thus a small number of short loops.  The 
\textit{Kesten-McKay law} \cite{Kesten} for the spectrum of sparse regular
graphs is a rare example of an analytical solution for the spectral density
and shows that regular graphs have a large spectral
gap, implying  many optimal structural
properties \cite{Murr}. Spectral analyzes of irregular sparse random graphs
such as Erd\"os-R\'enyi graphs \cite{Roger08, Bordenave}, scale-free graphs and
small-world
systems have recently been considered \cite{fortin}. 

However, Bravais lattices 
and real-world networks, such as the Internet
and metabolic networks, exhibit a large number
of undirected and directed short 
loops \cite{Gleiss}, while other examples
like power grids and neural networks are under-short 
looped, i.e.~they have less short loops than
their corresponding random graph models \cite{Bianc}.
To study the effect of loops on structural 
and dynamical properties of complex networks 
we consider the \textit{Husimi graph}
\cite{Husimi} (also called Husimi cactus), which is built out of randomly
drawn short loops.
The Husimi graph allows for a detailed spectral analysis 
as a function of the loop length, due
to its exactly solvable nature.
To our knowledge, results for the spectrum
of graphs with loops are scarce, apart from
the analytical formula for the triangular Husimi
graph \cite{Eck05}.

In this letter we present a systematic study of 
the spectra of regular Husimi graphs containing undirected
or directed edges, going beyond previous studies on local
tree-like networks without short loops.
We analyze the influence of loops on some important network properties:
the size of the 
spectral gap and the stability of
synchronized states.
The simplicity and exactness of our equations, confirmed
by direct diagonalization methods, leads to accurate results for
arbitrary loop lengths and allows 
for an extension of the Kesten-McKay law to triangular and square 
undirected Husimi graphs as well as to directed regular graphs
without short loops.
\begin{figure}[h!]
\begin{center}
\includegraphics[angle = 0, scale=0.3]{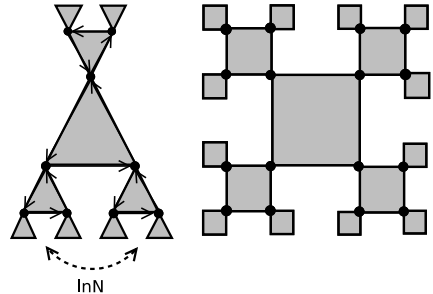}
\caption{Local tree-like structure of a $(3,2)$-directed and
a $(4,2)$-undirected regular Husimi graph. The average
path length between two nodes is of order $\mathcal{O}(\ln N)$.}
\label{fig:cavities}
\end{center}
\end{figure}  

\paragraph{Sparse regular graphs with loops}

We consider the ensemble of $(\ell,k)$-regular 
(un)directed Husimi graphs containing $N$ vertices or nodes.
Each vertex is incident to 
$k>1$ loops composed of 
$\ell$ nodes, with $k$ and $\ell$
independent of $N$. The indegree 
and outdegree of any node 
are equal to each other, and
given by $2k$ or $k$ in the case of 
undirected or directed 
Husimi graphs, respectively.
For $N \rightarrow \infty$ the graphs have a local tree-like 
structure on the level of loops, illustrated 
in figure \ref{fig:cavities} for 
triangular ($\ell = 3$) and square ($\ell = 4$)
Husimi graphs. The model allows to interpolate between 
$\ell = 2$ and $\ell \rightarrow \infty$, both 
cases representing situations 
where short loops are absent.
  
We study the spectral density of 
the $N \times N$ adjacency matrix $\bJ$ 
for $N \rightarrow \infty$, which is trivially
related to the spectrum of the Laplacian matrix in 
the case of regular graphs.  
The matrix element $J_{ij}$ assumes
$1$ if there is a directed edge from
node $i$ to  node $j$, and zero otherwise.
Denoting the eigenvalues of a given instance of $\bJ$
as $\{ \lambda_{i} \}_{i=1,\dots,N}$, the spectral density
is defined as
$\rho(\lambda) \equiv \lim_{N\rightarrow \infty}\frac{1}{N} \sum_{i=1}^{N}
\delta(\lambda- \lambda_{i})$.
The matrix $\bJ$ is symmetric or asymmetric
depending whether the graph is undirected or 
directed, respectively.
The eigenvalues are real in the former case
and complex in the latter.
The local tree-like structure shown in
figure \ref{fig:cavities} allows to calculate 
$\rho(\lambda)$ exactly for $N \rightarrow \infty$.

\paragraph{Spectra of undirected Husimi graphs}
 
The resolvent $\bG(z)$ of $\bJ$ is defined through 
$\bG(z) \equiv (z-\bJ)^{-1}$, where the complex
variable $z=\lambda - i \epsilon$ contains
a regularizer $\epsilon$. The spectrum 
is extracted from the diagonal
components of $\bG(z)$ according to
$\rho(\lambda) = \lim_{N\rightarrow \infty, \epsilon\rightarrow
0^{+}}(\pi N)^{-1} {\rm Im} {\rm Tr}
\bG(\lambda-i\epsilon)$.

Due to the absence of disorder, a closed expression can be derived for the 
diagonal elements $G_{ii}(z) = G(z),\,\, \forall \,\, i$. For graphs
without short loops, either one writes 
$G_{ii}(z) $ as the variance of a Gaussian function and uses the {\it cavity
method}
(or the replica method) \cite{Roger08}, or one uses repeatedly the
{\it Schur-complement
formula} and the local convergence of graphs to a tree \cite{Bordenave}.
Generalizing these methods to Husimi graphs \cite{MetzINP}, we
have derived the following equation
for $\rho(\lambda)$ 
\begin{equation}
\rho(\lambda) =  \frac{1}{\pi}  \lim_{\epsilon\rightarrow
0^{+}}  {\rm Im} [z-k \, G_{s}]^{-1} \,,
\label{eq:selfc1}
\end{equation}
where $G_{s}$ solves 
\begin{equation}
G_{s} = \mathbb{J}_{s}^{T} \Big[ \big( z - (k-1) G_{s} \big) \bI_{\ell-1}
- \bL_{\ell-1} - \bL_{\ell-1}^{T}    
\Big]^{-1} \mathbb{J}_{s} \,,
\label{cavsym}
\end{equation}
with $\bI_{\ell-1}$ the $(\ell-1) \times (\ell-1)$
identity matrix, $\bL_{\ell-1}$ the $(\ell-1)$-dimensional matrix
with elements $[\bL_{\ell-1}]_{i j} = \delta_{i,j-1}$, and
$\mathbb{J}_{s}^{T}$  the $(\ell-1)$-dimensional vector 
$\mathbb{J}_{s}^{T} = (1 \,\,\, 0 \,\, \dots \,\, 0 \,\,\, 1)$.
For $\ell=2$, the solution of eq.~(\ref{cavsym}) 
 yields the Kesten-McKay law \cite{Kesten}, where
$\rho(\lambda)$ takes the form
\begin{equation}
\rho(\lambda) = \frac{k}{2 \pi} 
\frac{\sqrt{4(k-1)-\lambda^{2}}}{k^{2} -\lambda^{2}}\,
\label{eq:Kesten:Mackay}
\end{equation}
for $|\lambda| < 2 \sqrt{k-1}$, and zero otherwise. 
For $\ell > 2$, we have inverted the matrix  
in eq.~(\ref{cavsym}) \cite{Huang97}, leading to
\begin{equation}
G_{s} = \frac{2 \alpha_{\ell-2} + 2}{\alpha_{\ell-1}}    
\,\,,
\label{GsA}
\end{equation}
where the coefficients $\alpha_{2},\dots \alpha_{\ell-1}$
follow from the recurrence relation
$\alpha_{i} = \alpha_{1} \alpha_{i-1} - \alpha_{i-2}$,
with $\alpha_{0} = 1$ and $\alpha_{1} =  z - (k-1) G_{s}$.
Equation (\ref{GsA}) leads to a polynomial 
in the variable $G_{s}$ and
can be solved analytically for smaller values of $\ell$, extending
the Kesten-McKay law to regular graphs containing
short loops. For larger values of $\ell$ a straightforward numerical
solution can be obtained, giving sharp 
results for $\rho(\lambda)$.
Equation (\ref{GsA}) is 
one of the main results of our 
work, allowing to compute exactly the spectrum 
for increasing values of $\ell$.

For $\ell=3$ we recover 
the analytical expression for $\rho(\lambda)$ 
presented in \cite{Eck05}. 
For $\ell=4$ eq. (\ref{GsA}) becomes a cubic polynomial 
with discriminant 
\begin{equation}
D(\lambda) = - \frac{2}{3}\lambda^{4} - \frac{\lambda^{2} }{3}
\left(k^2-22k+13  \right)
+ \frac{8}{3} (k-2)^{3}\,.
\end{equation}
Defining the functions
$s_{\pm}(\lambda) = 9\lambda (k+1) - \lambda^{3} \pm 9 \sqrt{D(\lambda)}\,$
and $q_{\pm}(\lambda) = s^{1/3}_{+} \pm s^{1/3}_{-}$, the 
spectrum of square Husimi graphs reads
\begin{equation}
\rho(\lambda) = \frac{6\sqrt{3} \, k \, (k-1) \, q_{-}(\lambda)}
{ \pi \Big[ 2 (k-3) \lambda + k \, q_{+}(\lambda)  \Big]^{2} 
+ 3 \, \pi \, k^{2} \, q_{-}^{2}(\lambda)
}
\end{equation}
for $D(\lambda) > 0$, and $\rho(\lambda) = 0$ otherwise. 
The edges of $\rho(\lambda)$ solve the equation
$D(\lambda) = 0$. The analytic expression for some higher 
values of $\ell$ is given elsewhere \cite{MetzINP}.

\begin{figure}[h!]
\begin{center}
\includegraphics[angle=0, 
scale = 0.28 ]{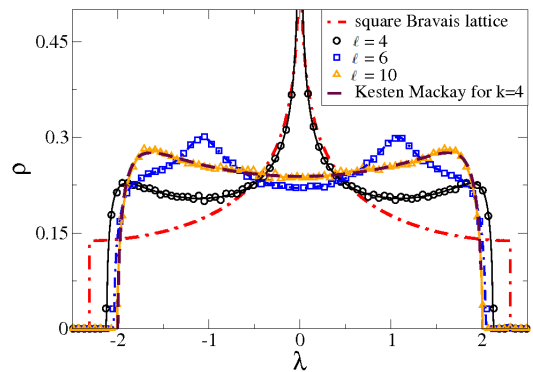}
\caption{Spectrum of $(\ell,k)$
undirected Husimi graphs with $k=2$ and 
$J_{ij} \rightarrow J_{ij}/\sqrt{2k-1}$, obtained by solving
eqs.~(\ref{eq:selfc1}) and (\ref{GsA}). The
symbols are direct
diagonalization results of adjacency matrices of
size $N = 10^{4}$. The spectrum of
the two-dimensional square Bravais lattice and 
the Kesten-McKay law are presented for
comparison.}
\label{fig:ellVar}
\end{center}
\end{figure}

In figure \ref{fig:ellVar} we compare direct diagonalization results 
of finite matrices  with the solution to 
eq.~(\ref{GsA}), for $k=2$ and several values of $\ell$. 
The agreement is excellent, following from
the exactness of eq.~(\ref{GsA}) for $N\rightarrow \infty$. When rescaling the
matrix elements $J_{ij} \rightarrow J_{ij}/\sqrt{2k-1}$ we 
find analytically the convergence of $\rho(\lambda)$ to the 
Wigner semi-circle law for $k \rightarrow \infty$ and 
arbitrary $\ell$ \cite{Bai}.
Interestingly, the 
spectrum of a square Husimi graph exhibits a striking
similarity with the spectrum of the two-dimensional square Bravais lattice \cite{Faz},
with the appearance of a power-law singularity at $\lambda=0$ with $\rho(\lambda)
\sim |\lambda|^{-1/3}$.
In the case of the square Bravais lattice, the spectral density
contains a Van Hove singularity at $\lambda=0$, with
a logarithmic divergence. Our results thus suggest that Van Hove
singularities
are related to the local
neighborhoods and not to the dimensional
nature of lattices \cite{Faz}.
For $\ell \rightarrow \infty$, the spectrum converges to
the Kesten-McKay law with degree $2k$ \cite{Kesten}, as
illustrated in figure \ref{fig:ellVar} for $\ell=10$.
Therefore, loops composed of ten nodes can be neglected and the
graph can be considered locally tree-like \cite{Roger08, Bordenave}.

\paragraph{Spectra of directed Husimi graphs}
  
In the case of directed Husimi graphs, the density 
of states $\rho(\lambda)$ at a certain point
$\lambda=x+ i y$ of the complex plane 
can be written as
$\rho(\lambda) = \lim_{N\rightarrow \infty}(N \pi)^{-1} \partial^{*} \text{Tr}
\bG(\lambda)$,
where $\partial^{*} = \frac{1}{2} \left(  \frac{\partial}{\partial x} + i  
\frac{\partial}{\partial y} \right)$ and $\bG(\lambda) = (\lambda - \bJ)^{-1}$.
The operation $(\cdot)^{*}$ denotes complex conjugation. 
Due to the non-analytic behavior of $G_{ii}(\lambda)$
in the complex plane \cite{FZ97}, it
is convenient to define the $2N \times 2N$ block matrix \cite{RPC09}
\begin{equation}
\bH_{\epsilon}(\lambda) =
\begin{pmatrix}
\epsilon \bI_{N} \hfill & -i(\lambda-\bJ) \\
-i(\lambda^{*}-\bJ^{T}) & \epsilon \bI_{N} \hfill 
\end{pmatrix} \,.
\label{defH}
\end{equation}
The $N \times N$ lower-left block of $\lim_{\epsilon \rightarrow
0^{+}} \bH^{-1}_{\epsilon}(\lambda)$
is precisely the matrix $\bG(\lambda)$. Thus,
the problem reduces to calculating the 
matrix elements $\mathcal{G}_{j}(\lambda, \epsilon) = 
\left[ \bH^{-1}_{\epsilon}(\lambda) \right]_{j+N,j}$ ($j=1,\dots,N$), 
from which the spectrum is determined according to 
$\rho(\lambda) = - \frac{i}{N \pi} \lim_{ N\rightarrow
\infty, \epsilon \rightarrow 0^{+}}
\sum_{j=1}^{N} 
\partial^{*} \mathcal{G}_{j}(\lambda,\epsilon)$.

By representing $\left[ \bH^{-1}_{\epsilon}(\lambda) \right]_{j+N,j}$ as
a Gaussian integral one can generalize the cavity 
method, as developed for sparse non-Hermitian random 
matrices \cite{RPC09}, to calculate the 
spectrum of directed Husimi graphs \cite{MetzINP}. 
Due to the absence of disorder we have that 
$\mathcal{G}_{j}(\lambda,\epsilon) = 
\mathcal{G}(\lambda,\epsilon), \,\, \forall \,\, j$, and $\rho(\lambda)$
is given by
\begin{equation}
\rho(\lambda) = \frac{1}{i \pi} \lim_{\epsilon \rightarrow 0} 
\partial^{*} \left[ \bS_{\epsilon}(\lambda)  + k \, \bG_{A}
\right]_{21}^{-1}\,,
\label{spectrcav}
\end{equation}
where $\bS_{\epsilon}(\lambda) =  \left[ \epsilon \bI_{2} - i 
\left(x \sigma_{x} - y \sigma_{y} \right)   \right]$
and $(\sigma_{x} , \sigma_{y})$ are Pauli matrices.
For $\ell > 2$, the two-dimensional matrix $\bG_{A}$ solves the equation
\begin{eqnarray}
\bG_{A} &= \mathbb{J}_{A}^{T} \Big [ \big( \bS_{\epsilon}(\lambda) + (k-1)
\bG_{A} \big) \otimes \bI_{l-1} \nonumber \\
&+ i \mathcal{J} \otimes \bL_{\ell-1} + i \mathcal{J}^{T} \otimes
\bL_{\ell-1}^{T}  \Big]^{-1} \mathbb{J}_{A}  \,,
\label{eqD}
\end{eqnarray}
where $\mathbb{J}_{A}^{T}$ is
the $2 \times 2(\ell-1)$ block matrix
$\mathbb{J}_{A}^{T} = ( \mathcal{J} \,\, 0 \, \dots \, 0 \,\, \mathcal{J}^{T})$,
with $\mathcal{J} = \frac{1}{2}(\sigma_{x} + i \sigma_{y})$.
The derivative of eq.~(\ref{eqD})
yields an equation in $\partial^{*}\bG_{A}$, which has to be 
solved together with (\ref{spectrcav}) to
find $\rho(\lambda)$. Equation (\ref{eqD}) 
allows to derive sharp numerical results
for the spectrum of directed Husimi graphs 
as a function of $\ell$.
\begin{figure}[h!]
\includegraphics[scale=1.1]{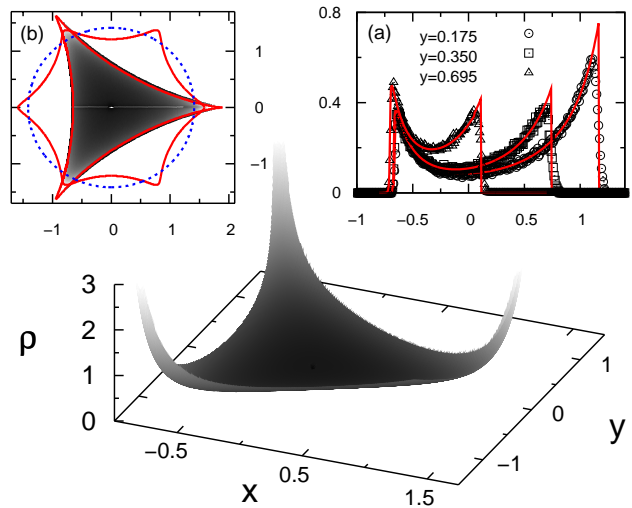}
\caption{Spectrum of directed Husimi 
graphs with $\ell=3$ and $k=2$, 
obtained from eqs.~(\ref{spectrcav}-\ref{eqD}). Inset (a) 
shows three cuts along the real direction (red 
curves), together with direct
diagonalization results (symbols) obtained
from an ensemble of $3 \times 10^{4}$ matrices
of size $N=10^{3}$.  Inset (b) shows
theoretical results for the boundary 
of $\rho(\lambda)$ for $\ell=3$ and 
$\ell=6$ (red curves).
The number of corners in each boundary 
is equal to the value of $\ell$ and the
blue dashed curve corresponds to the circle 
$|\lambda|^{2}=k$, for $\ell\rightarrow \infty$. For comparison,
direct diagonalization
results are also shown in grey scale for
$\ell=3$.
}
\label{CompDiag}
\end{figure}   

In figure \ref{CompDiag} we present the spectrum $\rho(\lambda)$ for
$\ell=3$ and $k=2$, comparing the 
solution to eqs.~(\ref{spectrcav}-\ref{eqD})
with direct diagonalization results. 
The agreement is excellent. A prominent feature of 
$\rho(\lambda)$ is the $\ell$-fold rotational 
symmetry, due to the transformation properties of 
$\bG_{A}$ under rotations of $2 \pi/\ell$. 
By rescaling 
$J_{ij} \rightarrow J_{ij}/\sqrt{k-1}$, we find analytically the convergence
of $\rho(\lambda)$ to Girko's circular law for $k \rightarrow \infty$ and
arbitrary $\ell$ \cite{Bai}. 

Analogously to undirected
Husimi graphs, $\rho(\lambda)$ converges to the spectrum
of a directed regular graph without short loops for $\ell \rightarrow \infty$.
In this case, we find a remarkable extension 
of the Kesten-McKay law, Eq. (\ref{eq:Kesten:Mackay}), to directed graphs,
where $\rho(\lambda)$ takes the form
\begin{equation}
\rho(\lambda)= \frac{k-1}{\pi}  \left( \frac{k }{k^2 -|\lambda|^{2}}
\right)^{2}\,,
\label{spectruml2}
\end{equation}
for $|\lambda|^{2} < k $, and zero otherwise.
A comparable equation appeared in \cite{R10}, but with a 
different support of $\rho(\lambda)$.

In inset (b) of figure \ref{CompDiag} we plot 
the boundary of $\rho(\lambda)$ for $k=2$ and increasing values 
of $\ell$. In accordance with eq.~(\ref{spectruml2}), the
boundary converges to the circle $|\lambda|^{2} = k$
in the limit $l \rightarrow \infty$. For $\ell=10$ we have
obtained numerically that $\rho(\lambda)$ is given
approximately by 
eq.~(\ref{spectruml2}) and the graph becomes 
locally tree-like \cite{RPC09}.

\paragraph{Structural and dynamical properties}

Let us order the eigenvalues of a regular 
undirected Husimi graph as
$\lambda_1<\lambda_2<\cdots < \lambda_N$, where $\lambda_N =
2k$. The \textit{spectral gap} $g$ and the  \textit{eigenratio} $Q$ are,
respectively, defined by $g \equiv
(\lambda_N-\lambda_{N-1})/2k$ and $Q \equiv
(\lambda_N-\lambda_1)/(\lambda_N-\lambda_{N-1})$. 
Analogously, for regular directed Husimi graphs, the eigenvalues can be ordered
according to their real parts ${\rm Re} \lambda_1< {\rm Re} 
\lambda_2<\cdots < {\rm Re} \lambda_N$, with 
${\rm Re} \lambda_N = k$. In this case, the spectral
gap $g$ and the eigenratio $Q$ are given by $g \equiv ({\rm Re} \lambda_N - {\rm
Re} \lambda_{N-1} )/k$ 
and $Q \equiv ({\rm Re} \lambda_N- {\rm Re}\lambda_1)/({\rm Re}\lambda_N-{\rm
Re}\lambda_{N-1})$.
 
The spectral gap $g$ controls the speed
of convergence to the stationary state of 
diffusion processes on the graph \cite{Bar}. 
Designing communication
networks with a large
$g$ is known to be important due to improved robustness
and communication properties \cite{Murr, Hoory}, for undirected networks. 
The eigenratio $Q$ measures the 
propensity for synchronization 
in networks of oscillators \cite{Pec,Don}.
A linear stability analysis shows that 
synchronized states are more stable
for smaller values of $Q$.

\begin{figure}[h!]
\includegraphics[angle=0, 
scale=0.27]{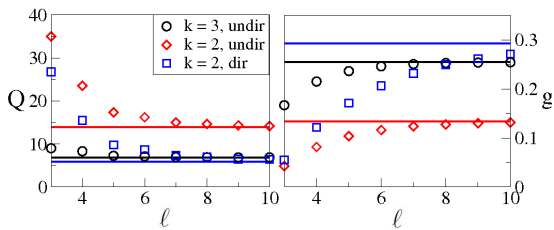}
\caption{Spectral gap $g$ and eigenratio $Q$ of
Husimi graphs as functions of $\ell$ for different
values of $k$, with the asymptotic behavior
for $\ell \rightarrow \infty$ indicated by solid lines.}
\label{spectralGap}
\end{figure}

Figure \ref{spectralGap} depicts $g$ and $Q$ as
functions of $\ell$ for regular Husimi graphs, showing
that $g$ increases while $Q$
decreases for increasing values of $\ell$.
For undirected Husimi graphs, $g$ and $Q$
converge, respectively, to $(k-\sqrt{2k-1})/k$ 
and $2k/(k-\sqrt{2k-1})$ as $\ell 
\rightarrow \infty$, consistent with the Alon-Boppana 
bound for the second largest eigenvalue \cite{AB}.
For directed Husimi graphs
$g$ and $Q$ converge to $(k-\sqrt{k})/k$
and $2k/(k-\sqrt{k})$, respectively.
In summary, short loops
have a negative influence on the
synchronization properties and
on the size of the spectral gap, which is more pronounced 
at low connectivities.

\paragraph{Conclusions}

We have determined the spectrum of sparse regular 
random graphs with short loops through 
a set of exact equations, including extensions
of the Kesten-McKay law to triangular and square
undirected Husimi graphs as well as to directed
regular graphs without short loops. 
We find that short loops in directed and undirected
networks have a negative influence on the stability
of synchronized states, they also worsen the communication properties due to a
decrease of the spectral gap.  
Our spectral results make the absence of loops
in network construction apparent \cite{Don}, while 
neural networks are under-short looped \cite{Bianc}.
For the square Husimi graph we recover a singularity at the
origin, which is also present in a square Bravais lattice. 
Overall, we find that the spectra of Bravais lattices
are similar to the spectra of Husimi graphs with 
suitable neighborhoods, indicating that Husimi 
graphs serve as good toy models for Bravais lattices.
Our results on spectra of sparse
random matrices are of wide
interest to diverse fields including the study of Markov chains \cite{Norris},
dynamics of spin-glasses \cite{Semerj}, etc. Since our
work is mainly based upon the cavity method, it 
allows for an extension to e.g.~irregular
graphs with loops \cite{Roger08} and eigenvector localization studies
\cite{Abou73}.

FLM thanks Reimer K\"uhn and Isaac P\'erez
Castillo for interesting discussions, and
Tim Rogers for a useful correspondence.


\vspace{-1cm}             
\begin{thebibliography}{99}
\bibitem{Bar}A.\,Barrat, M.\,Barth\'elemy, A.\,Vespignani, \textit{Dynamical
processes on networks}, Cambridge University Press (2008); M.\,E.\,J.\,Newman,
\textit{Networks, An Introduction}, Oxford University Press (2010).
\bibitem{Hoory}S.\,Hoory, N.\,Linial, A.\,Wigderson, Bull. Amer. Math. Soc.
{\bf 43}, 439 (2006).
\bibitem{Murr}M.\,R.\,Murty, J. Ramanujan Math. Soc. {\bf 18}, 1 (2003).
\bibitem{Pec}M.\,Barahona, L.\,M.\,Pecora, Phys. Rev. Lett. {\bf 89}, 054101
(2002).
\bibitem{Don}L.\,Donetti, P.\,I.\,Hurtado, M.\,A.\,Mu\~{n}oz, Phys. Rev.
Lett. {\bf 95}, 188701 (2005);T.\,Nishikawa, A.\,E.\,Motter, Phys. Rev. E {\bf
73}, 065106 (2006).
\bibitem{Lov}L.\,Lov\'asz, P.\,Winkler, \textit{Mixing of random walks and other
diffusions on a graph}, Cambridge University Press (1995), 119-154.
\bibitem{Chung}F.\,R.\,K.\,Chung, {\it Spectral Graph Theory}, American
Mathematical Society (1997);
B.\,Georgeot, O.\,Giraud, D.\,L.\,Shepelyansky, Phys. Rev.
E {\bf 81}, 056109 (2010); Z.\,Wang, A.\,Scaglione, R.\,Thomas, IEEE
Transactions on Smart Grid {\bf 1}, 1949-3053 (2010). 
\bibitem{Bol}B.\,Bollob\'as, {\it Random graphs}, Cambridge University Press
(2001).
\bibitem{Kesten}H.\,Kesten, Trans. Amer. Math. Soc. {\bf 92}, 336354 (1959); 
B.\,D.\,McKay, Linear Algebra Appl. {\bf 40}, 203-216 (1981).
\bibitem{Roger08}T.\,Rogers, K.\,Takeda, I.\,P.\,Castillo and R.\,K\"uhn, 
Phys. Rev. E. {\bf 78}, 031116 (2008);  R.\,K\"uhn, J. Phys. A: Math. Theor.
{\bf 41}, 295002 (2008).
\bibitem{Bordenave}C.\,Bordenave, M.\,Lelarge, Random Structures and Algorithms
{\bf 37}, 332 (2010).
\bibitem{fortin}J.\,Fortin, J. Phys. A: Math. Gen. {\bf 38}, L57 
(2005); A.\,N.\,Samukhin, S.\,N.\,Dorogovtsev, J.\,F.\,F.\,Mendes,
Phys. Rev. E {\bf 77}, 036115 (2008); I.\,J.\,Farkas, 
I.\,Der\'enyi, A.\,-L.\,Barab\'asi and T.\,Vicsek, Phys. Rev. E, {\bf
64}, 026704 (2001); 
R.\,K\"uhn, J.\,Mourik, J. Phys. A: Math. Theor. {\bf 44}, 165205 (2011).
\bibitem{Gleiss}P.\,M.\,Gleiss et al., arxiv:cond-mat/0009124 
(2000); G.\,Bianconi, G.\,Caldarelli, A.\,Capocci, Phys. Rev. E {\bf 71}, 066116
(2005).
\bibitem{Bianc}G.\,Bianconi, N.\,Gulbahce, A.\,E.\,Motter, Phys. Rev. Lett. {\bf
100}, 118701 (2008).
\bibitem{Husimi}K.\,Husimi, J. Chem. Phys. {\bf 18}, 682-684
(1950);  F.\,Harary, G.\,Uhlenbeck, PNAS {\bf 39}, 315-322 (1953).
\bibitem{Eck05} M.\,Eckstein, M.\,Kollar, K.\,Byczuk and
D.\,Vollhardt, Phys. Rev. B {\bf 71}, 235119 (2005); M.\,Galiceanu, A.\,Blumen, J. Chem. Phys.
{\bf 127}, 1349004 (2007).
\bibitem{MetzINP} F.\,L.\,Metz, I.\,Neri, D.\,Boll\'e, in preparation.
\bibitem{Bai}Z.\,D.\,Bai, J.\,W.\,Silverstein, {\it Spectral Analysis of Large
Dimensional Random Matrices}, Science Press (2006).
 \bibitem{Huang97} Y.\,Huang, W.\,F.\,McColl, J. Phys A: Math. Gen. {\bf 30},
7919 (1997).
\bibitem{Faz}P.\,Fazekas, \textit{Lecture notes on Electron Correlation and
Magnetism}, World Scientific (1999).
\bibitem{FZ97} J.\,Feinberg, A.\,Zee, Nucl. Phys. B {\bf 504}, 579 (1997).
\bibitem{RPC09} T.\,Rogers and I.\,P.\,Castillo, Phys. Rev. E {\bf 79},
012101 (2009).
\bibitem{R10} T.\,Rogers, \textit{New results on the spectral density of random
matrices}, thesis (2010). 
\bibitem{AB}N.\,Alon, Combinatorica {\bf 6}, 83 (1986); J.\,Friedman, Mem. Amer. Math. Soc. {\bf 195}, 910 (2008).
\bibitem{Norris}J.\,R.\,Norris, {\it Markov Chains, Statistical and
Probabilistic
Mathematics}, Cambridge University Press (1998).
\bibitem{Semerj} G.\,Semerjian and L.\,F.\,Cugliandolo, Europhys. 
Lett. {\bf 61}, 247 (2003).
\bibitem{Abou73} G.\,Biroli, G.\,Semerjian, M.\,Tarzia, Prog.
Theor. Phys. Suppl. {\bf 184}, 187 (2010); F.\,L.\,Metz, I.\,Neri, D.\,Boll\'e,
Phys. Rev. E {\bf 82}, 031135 (2010).
\end{thebibliography}
\end{document}